\documentclass[sigconf,screen]{acmart}

\AtBeginDocument{%
  }

\setcopyright{acmlicensed}
\acmDOI{10.1145/3663529.3663788}
\acmYear{2024}
\copyrightyear{2024}
\acmSubmissionID{fsecomp24ivr-p93-p}
\acmISBN{979-8-4007-0658-5/24/07}
\acmConference[FSE Companion '24]{Companion Proceedings of the 32nd ACM International Conference on the Foundations of Software Engineering}{July 15--19, 2024}{Porto de Galinhas, Brazil}
\acmBooktitle{Companion Proceedings of the 32nd ACM International Conference on the Foundations of Software Engineering (FSE Companion '24), July 15--19, 2024, Porto de Galinhas, Brazil}
\received{2024-01-29}
\received[accepted]{2024-04-09}

\usepackage[]{mdframed}

\begin{document}

\title{A Vision on Open Science for the Evolution of Software Engineering Research and Practice}

\author{Edson OliveiraJr}
\affiliation{
   \institution{State University of Maringá}
   \city{Maringá}
   \country{Brazil}}
\email{edson@din.uem.br}

\author{Fernanda Madeiral}
\affiliation{
   \institution{Vrije Universiteit Amsterdam}
   \city{Amsterdam}
   \country{Netherlands}}
\email{fer.madeiral@gmail.com}

\author{Alcemir Rodrigues Santos}
\affiliation{
   \institution{State University of Piauí}
   \city{Piripiri}
   \country{Brazil}}
\email{alcemir@prp.uespi.br}

\author{Christina von Flach}
\affiliation{
   \institution{Federal University of Bahia}
   \city{Salvador}
   \country{Brazil}}
\email{flach@ufba.br}

\author{Sergio Soares}
\affiliation{
   \institution{Federal University of Pernambuco}
   \city{Recife}
   \country{Brazil}}
\email{scbs@cin.ufpe.br}

\begin{abstract}
Open Science aims to foster openness and collaboration in research, leading to more significant scientific and social impact. However, practicing Open Science comes with several challenges and is currently not properly rewarded. In this paper, we share our vision for addressing those challenges through a conceptual framework that connects essential building blocks for a change in the Software Engineering community, both culturally and technically. The idea behind this framework is that Open Science is treated as a first-class requirement for better Software Engineering research, practice, recognition, and relevant social impact. There is a long road for us, as a community, to truly embrace and gain from the benefits of Open Science. Nevertheless, we shed light on the directions for promoting the necessary culture shift and empowering the Software Engineering community.
\end{abstract}

\ccsdesc[500]{Software and its engineering}

\keywords{Open Science, artifacts, research software, education, culture shift}

\maketitle

\section{Introduction}

In the 21st century, the \textit{Open Science} movement is a significant paradigmatic transformation entailing a systemic change in science and research in all disciplines. It designates the practice of science in a way that \textit{(i)} others can collaborate and contribute, and \textit{(ii)} publications, data, software, and other research artifacts are made available, reusable, and reproducible~\citep{unesco:2021,munafo2017manifesto}. In addition, Open Science practices combat fraud and improve research rigorousness~\citep{Seibold2024}.

However, as many significant transformations and innovations in the history of science had to deal with multiple challenges and barriers~\citep{nas2018}, Open Science benefits also come with their own. \citet{Bartling2014} defined the transition towards Open Science as a complex cultural change, and \citet{Guzzo:IOP:2022} stressed some incompatibilities between Open Science principles and scientific progress through applied research. This suggests no straightforward recipe for embracing Open Science, but a series of actions should be taken.


In Software Engineering, there has been some discussion about Open Science. For instance, 
\citet{mendez2020} discuss Open Science practices, \textit{e.g.}, open materials and registered reports, and how the Empirical Software Engineering community can adopt them.
With regard to education and training, researchers and representatives for research laboratories and funding agencies have joined meetings and workshops to discuss the need for Research Software Engineers and possible actions~\citep{rs:et2022,rs:wosss,KatzHettrick2023}. 


Although preliminary discussion exists in the literature, it still lacks an answer to the question: ``\textit{How can our Software Engineering community adopt the Open Science movement?}''~\citep{mendez2020}. 
Researchers can no longer ignore or treat Open Science issues as incidental but adopt them as a \textit{first-class research requirement}. For instance, several universities are currently restructuring their research career assessment process, shifting careers from a paper-based model to a social-contribution-based one~\citep{Chawla2021}. Such a change comes with Open Science principles and practices attached~\citep{nas2018}, and Software Engineering researchers will increasingly need to work their way toward putting those into practice.
Accordingly, research artifacts are present in any research project and should, therefore, be recognized, evaluated, and published together with the associated papers for the sake of Open Science. 

In this paper, we share our reflections and visions about Open Science, with a focus on \textit{research artifacts}, as an attempt to structure future research to answer \citet{mendez2020}'s question. Artifacts are the backbone of research projects and, consequently, of research papers. Although creating and sharing reusable research artifacts is challenging, they allow transparency, replications, and reuse in new research projects, making it possible for a research community to mature. We present our vision as a framework relating actions that should be taken toward Open Science in Software Engineering.

\section{Our Vision}

\begin{figure}[!h]
    \hspace*{-4pt}
    \includegraphics[width=0.488\textwidth]{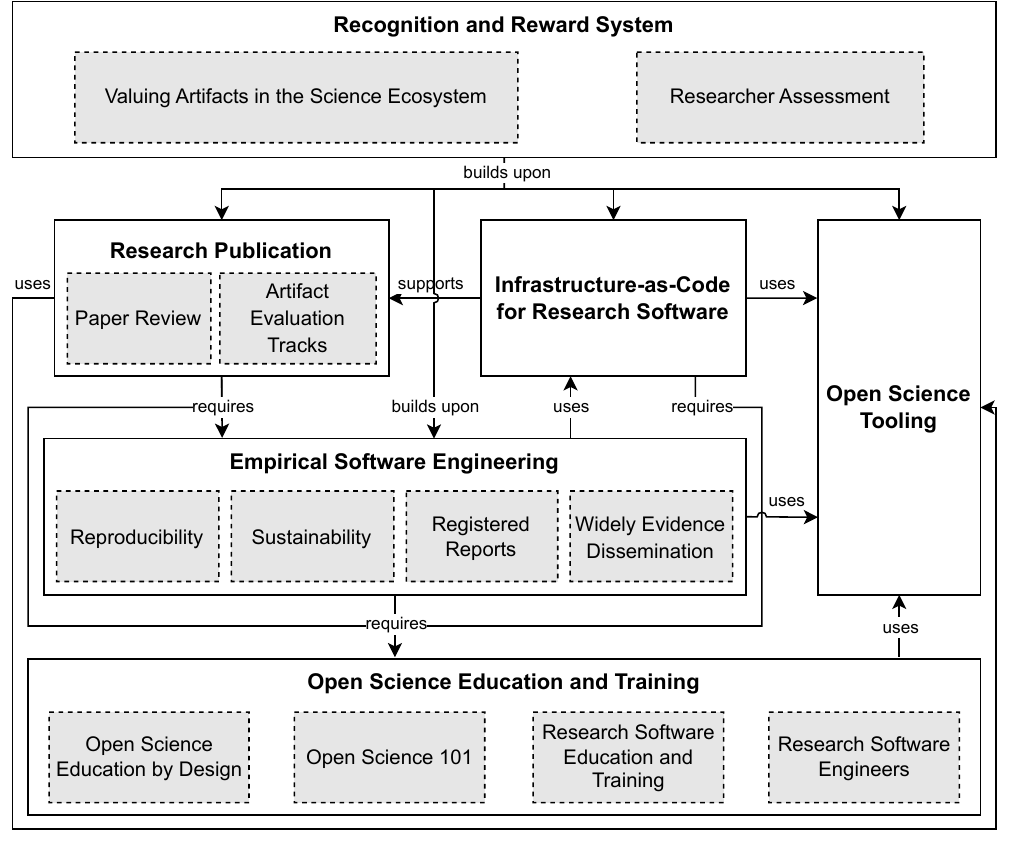}
    \vspace{-12pt}
    \caption{Our vision on Open Science for Software Engineering research and practice.}
    \label{fig:vision}
    \vspace{-6pt}
\end{figure}

\autoref{fig:vision} shows an overview of our vision for addressing the challenges in adopting Open Science practices, where building blocks that support each other are connected. First, the modern Software Engineering research career has to rely on Open Science principles and practices in its \textit{recognition and reward system}~\cite{howison2011scientific}. Research artifacts should be considered as a part of \textit{research publication}, just like papers are. Moreover, research software should be accompanied by \textit{infrastructure-as-code} to improve the \textit{Empirical Software Engineering} practice, as well as to support artifact evaluation during paper or artifact review. To realize those, we need to address \textit{Open Science education and training} so that the research community follows Open Science principles by design~\citep{nas2018} in their research. Finally, \textit{Open Science tooling} is required to support researchers with several tasks in several contexts. We elaborate on each building block in the following sections.

\subsection{Recognition and Reward System}

Published papers and citations are the coins to pay for the researchers' work~\citep{howison2011scientific}. However, source code, curated datasets, evidence briefings, and research software all have value in the construction of Software Engineering knowledge. Preparing and sharing them is challenging and time-consuming, which might be seen as not worth it if such an effort is not properly recognized and rewarded. Therefore, we need more space to produce, cite, and get recognition from different types of research artifacts. We argue that such recognition should go beyond paper badging through Artifact Evaluation Tracks, which is how researchers get some reward for producing and sharing good artifacts.
In our vision, artifacts should be valued in the science ecosystem, and researchers should be assessed by taking their Open Science practices into account.

\subsubsection{Valuing Artifacts in the Science Ecosystem}

Research software exists in many kinds of research due to the need for automated data analyses. In Software Engineering, it is also often the case of prototypes that serve as proof for proposed approaches that would ideally be used by developers in the industry. Sharing them is important for scientific advancement because several research projects rely on research software from previous work. However, they end up hidden in the results of a research publication. This is because research papers are the ultimate concrete object considered as the result of a research project. Research software and other artifacts, on the other hand, are not.
Because of their importance, beyond software citation~\citep{katz:citation}, there should exist additional mechanisms for recognizing and valuing research software and other artifacts in the science ecosystem, \textit{e.g.}, a Google Scholar for research software, other artifacts, and open reviews with 
blockchain-based token rewards~\citep{Santos:OpenScienSE:2021}.

\subsubsection{Researcher Assessment}

Any assessment of researchers, \textit{e.g.}, Ph.D. evaluation for obtaining the degree, hiring processes, and career assessment for promotions~\citep{European2017}, considers publications but rarely research artifacts.
Preparing and sharing research software is challenging and time-consuming, and researchers might neglect it due to a lack of recognition and reward.
While researcher assessment is mainly done over publications, Open Science practices have started to be required by funding agencies, \textit{e.g.}, in the Netherlands~\citep{Chawla2021}. This is due to initiatives for more research openness and sustainable software, which should be expanded in all events assessing researchers. 
The adoption of Open Science principles and practices in Software Engineering research may change incentive and reward systems~\cite{howison2011scientific,howison2013incentives}, promoting balanced recognition of papers and research artifacts.


\subsection{Research Publication}

Research papers are the main, and sometimes only, concrete outcome of a research project. This is how research has been published and disseminated. However, research artifacts are part of any research project and, therefore, should be published together with the associated papers for the sake of Open Science.

\subsubsection{Paper Review}

Researchers submit their papers to journals and conferences, which are evaluated based on a series of criteria. These criteria usually include relevance, novelty, soundness, and clarity. However, compliance with Open Science policies is often neglected. 
Even though some conferences have defined evaluation criteria for verifiability and transparency, artifact evaluation is still optional.  
For instance, the call for papers for the ICSE 2024 research track includes the following criteria: ``\textit{Verifiability and Transparency: The extent to which the paper includes sufficient information to understand how an innovation works; to understand how data was obtained, analyzed, and interpreted; and how the paper supports independent verification or replication of the paper’s claimed contributions. Any artifacts attached to or linked from the paper may be checked by one reviewer}''. 
Moreover, although one reviewer may check artifacts, acceptance is not conditional to artifact sharing. 
In our vision, artifact sharing, or a compelling justification for not doing so, should be a requirement for acceptance, independent of other reviewing criteria.
If not enforced, changing the current practice will be difficult because researchers will naturally prefer to focus their time on other activities than on an optional task.

\subsubsection{Artifact Evaluation Tracks}\label{sec:Research-Publication:Artifact-Evaluation-Tracks}

Artifact Evaluation Tracks call for submissions of artifacts produced for already accepted papers. The calls include a set of badges that are supposed to stamp papers indicating the availability or quality of the artifacts related to the papers. Researchers submit artifacts claiming for specific badges, which are evaluated according to the reward criteria. Even though paper badging is the current way researchers are rewarded for producing and sharing good research artifacts, the existence of Artifact Evaluation Tracks also means that creating and sharing reusable artifacts is not the default expected outcome from researchers. This does not comply with the idea of the Open Science movement. Therefore, artifact evaluation should be either directly incorporated in the paper reviewing process (when the community has the proper resources for that, see \autoref{sec:iac} and \autoref{sec:OpenScienceTooling}), or should exist only for advanced recognition (\textit{e.g.}, for the ACM ``Results Validated'' badges~\citep{ACMBadges}).


\subsection{Empirical Software Engineering}

Empirical Software Engineering has played a central role in the last decades as an effective way to provide reliable evidence for various research topics. However, such studies have not practically evolved concerning reproducibility, transparency, and openness~\citep{Gonzalez-BarahonaRobles2023}. Open Science practices and principles might contribute to effectively evolving Software Engineering empirical studies in the following four perspectives.

\subsubsection{Reproducibility}

The reproducibility of empirical studies might be reached by defining open data life cycles and providing a detailed description of artifact provenance, preservation, and curation. Such Open Science characteristics make empirical studies able to be audited and reproducible, thus allowing one to understand the whole data and artifacts cycle, provenance, and prospective traceability of evidence~\citep{Chirigati:SIGMOD:2016, Rampin:2018:reproserver,Pimentel:VLDB:2017,Cordeiro2022, CordeiroOliveiraJr2021,OliveiraJr_et_al2021}. 
Open replication packages also contribute to reproducibility. They should include all analyzed data (open data), research software (\textit{e.g.}, scripts), workflows, and other research artifacts necessary to comprehend and run the study.
In addition, they should follow the FAIR~\citep{WilkinsonEtAl2016} principles for findable, accessible, interoperable, and reusable data. 
Provenance techniques must support such packages to guarantee their traceability over time, derivation, and manipulation techniques that might affect previous and prospective studies and curation procedures.


\subsubsection{Sustainability}

As the research evolves, new replication packages are made available, thus demanding preservation techniques aimed at keeping data and other artifacts conserved and maintained. In addition, such techniques must consider policies for Green Computing~\citep{Paul_et_al2023} most related to energy consumption to the long-term keeping of artifacts deposited in repositories.

\subsubsection{Registered Reports}

Registered reports are essential for rigorous empirical research. They contribute to avoiding misconduct and planning to deviate by establishing a peer-reviewed protocol before an empirical study is conducted~\citep{nas2018}. 
Although they are little explored in the Software Engineering literature~\citep{NeilBaldassarre2023}, they represent a way of putting Open Science into practice by, for instance, demanding them for artifacts provenance and preservation, encompassing a Data Management Plan (DMP) for peer reviewing of such report, thus a fair operation. 

\subsubsection{Widely Evidence Dissemination}

Most empirical studies in Software Engineering lack proper dissemination of their findings, especially their artifacts. Finding documented study protocols or even their data for reproducibility is not trivial. 
Therefore, open empirical portals might effectively aid researchers and practitioners in localizing such artifacts. Such portals must encompass formal descriptions of empirical studies from ontologies, for instance, with proper metadata, as well as implementing features to facilitate users to perform Open Science practices such as provenance and curation~\citep{Cordeiro2022}. These portals might integrate with open repositories such as Zenodo or FigShare via APIs to deposit large artifacts and receive a unique identification as a DOI~\citep{Santana_et_al2023}.


\subsection{Infrastructure-as-Code for Research Software}\label{sec:iac}

Research software supports studies reported in scientific papers~\citep{Cohen:IEEESoftware:2020}. Several kinds of studies that build upon previous research, \textit{e.g.}, replications and empirical comparisons, would require access to research software produced in previous research, as well as clear and complete documentation to properly execute it. However, sharing reusable research software is challenging due to its dynamic nature~\citep{wiese:2020}. 
Sharing also involves knowledge about several steps to execute and operate software, such as installation and configuration, a challenge also faced by professionals who develop or operate software systems in the industry.

Even if authors can make research software available with proper documentation, it might be challenging for researchers to follow all the instructions to execute it and reproduce results. As with replications and reuse of software in new research projects, artifact evaluation also requires executing research software, depending on the artifact kind. This comes with additional challenges, including the lack of knowledge from reviewers on the research software, the potential need for familiarity with the research topic it belongs to, and limited time to read a full paper and execute many steps to verify if the software produces what it should. 

In our vision, researchers should make research software available accompanied by infrastructure-as-code. 
Infrastructure-as-code is the practice of preparing an infrastructure for running a software system by means of executable code. In such a format, whoever runs the software system would not need to read and understand documentation and then manually create and prepare the infrastructure. Instead, a script would do this task, allowing the successful execution of research software in an easier way. In combination with automated pipelines, if research software is made available with infrastructure-as-code, one could execute one single command that would result in a table or figure reported in a paper, for instance. Such a feature would make reproducing results and evaluating artifacts so easy that paper-related tracks could incorporate Artifact Evaluation as a criterion for accepting a paper.


\subsection{Open Science Education and Training}


A short-term concern for Software Engineering research and practice is the need for education and training on Open Science and Research Software. This involves the development, adaptation, or systematization of the openness definition~\citep{Levin_et_al2016}, open methodologies, courses, and workshops mainly based on Open Educational Resources (\textit{e.g.}, syllabi, materials, 
web portals, and MOOCs).

\subsubsection{Open Science Education by Design}

Open Science by Design refers to principles and practices to foster openness throughout the entire research life cycle by making research artifacts associated with their publications openly available under the FAIR principles~\citep{nas2018}.
%
We envision Open Science \textit{Education} by Design as a set of strategies for introducing Open Science principles and practices 
in formal education as early as possible.
In Software Engineering Education, potential strategies may address them in one discipline or as topics that crosscut several disciplines.
Students' comprehension 
can be assessed by direct application in an authentic research context, such as graduate and undergraduate theses.

\subsubsection{Open Science 101}\label{subsub:os101}

Open Science may be introduced in Software Engineering curricula as \textit{Open Science 101}, an introductory course addressing core Open Science concepts, the fundamentals of methods, practices, and tools for reproducible research~\citep{Opensciency2023}, and recommending 
them for undergraduate research projects.
An update to the Software Engineering curriculum guidelines~\citep{securriculum2014}, for instance, could include these topics in a new knowledge unit called ``\textit{Open Science Foundations for Software}''.

\subsubsection{Research Software Education and Training}

Scientists often develop domain software~\citep{hettrick2014uk} to support their research, and reproducibility requires it to be open and sustainable. Still, researchers need to be aware of Software Engineering best practices, including licenses, version control, coding standards, code documentation, testing, and release engineering~\citep{FOSTER, TheCarpentries}. Basic training on such practices is vital for fostering reproducible research.
Alternatively, research software engineers can provide professional support concerning research software to domain scientists.

\subsubsection{Research Software Engineers}

A research software engineer combines professional Software Engineering expertise with an understanding of research~\citep{rse:org} to work as an integral member of research groups or institutions, addressing the challenges of developing high-quality complex and customized software systems under Open Science principles for various scientific domains
\citep{rse:princ, rse:shef}.

We envision at least two practical approaches to supporting such knowledge and skills: a 4-year Software Engineering undergraduate degree with specialized or intensive research-focused courses at the 4th year level (honors degree) and new graduate programs (\textit{e.g.}, Masters in Research Software) with fundamentals from \textit{Open Science 101} 
and new knowledge units. These should include
(1)~Open and FAIR Research Data; 
(2)~Open, FAIR, and Sustainable Research Software;
(3)~Software Engineering for Research Software (quality attributes, open licenses, documentation, life cycle models, development and maintenance, testing and analysis, and CI/CD);
(4)~Recognition, Curation, and Preservation of Research Software;
(5)~Infrastructure-as-Code for Research Software; 
(6)~HPC and AI applications for large and complex tasks; and
(7)~Open Science tools.


\subsection{Open Science Tooling}\label{sec:OpenScienceTooling}

One challenge in practicing Open Science is the scarcity of tools to aid researchers in producing, assessing, and making research artifacts available and reusable in compliance with Open Science principles. For instance, FAIR guiding principles for Data~\citep{WilkinsonEtAl2016} and Research Software~\citep{chue_hong_fair_2022} help researchers think about their research artifacts and how to make them available and reusable. However, tool support is needed for researchers and artifact reviewers to assess how compliant such artifacts are with Open Science and FAIR principles.
One well-known challenge is to deal with sensitive data. 
Researchers may not share sensitive data because preparing the data to avoid leaking confidential information is difficult and time-consuming. Tools for aiding researchers in handling such cases would foster data sharing.

Our vision is that more specific tooling is needed to help researchers with several tasks, considering several conditions, for the openness and assessment of research artifacts. One such example is Anonymous GitHub~\citep{AnonymousGithub}, proposed and extensively used for anonymizing GitHub repositories, allowing researchers to share their artifacts during the double-anonymous review process.
Other useful tools include the set of tools of the OpenAire initiative for promoting Open Science, such as EOSC and FAIRCORE4EOSC, the European Open Science Cloud~\citep{OpenAire2024}, 
Reprozip~\citep{Chirigati:SIGMOD:2016}, 
ReproServer~\citep{Rampin:2018:reproserver}, and 
NoWorkflow~\citep{Pimentel:VLDB:2017}.


\section{Final Remarks}


To situate the adoption of Open Science in Software Engineering research and practice, we use an analogy to the adoption of Empirical Software Engineering practices 20 years ago.
Empirical Software Engineering has played a central role in the last decades as an effective way to provide reliable evidence for various research topics. 
However, its broad adoption began when Software Engineering venues enforced its principles and evaluation methods in their calls for papers and review processes, and graduate programs promoted education in Empirical Software Engineering. 
In our vision, a similar culture shift is underway in the context of Open Science adoption in Software Engineering research and practice.
This adoption requires efforts on Open Science education and training, valuing openness in Empirical Software Engineering, better infrastructures and tools to support Open Science practices, Open Science-aware paper reviewing and artifact evaluation tracks, and changes in the current recognition and reward systems. 

The Software Engineering community should not treat Open Science as incidental and start seamlessly adopting its principles and practices, for instance, while writing papers, sharing research artifacts, rewarding artifacts and papers that share them, or evaluating a CV for a research position.









\begin{acks}
Edson OliveiraJr thanks CNPq grant \#311503/2022-5. Fernanda Madeiral is partially supported by the European Commission grant 101120393 (Sec4AI4Sec) and the Dutch Research Council (NWO). Sergio Soares is partially supported by CNPq grant 306000/2022-9. This work is partially supported by INES 2.0 (www.ines.org.br), CNPq grant 465614/2014-0, FACEPE grants APQ-0399-1.03/17 and APQ/0388-1.03/14, CAPES grant 88887.136410/2017-00.
\end{acks}

\balance

\bibliographystyle{ACM-Reference-Format}
\bibliography{references}


\begin{thebibliography}{42}


\ifx \showCODEN    \undefined \def \showCODEN     #1{\unskip}     \fi
\ifx \showDOI      \undefined \def \showDOI       #1{#1}\fi
\ifx \showISBNx    \undefined \def \showISBNx     #1{\unskip}     \fi
\ifx \showISBNxiii \undefined \def \showISBNxiii  #1{\unskip}     \fi
\ifx \showISSN     \undefined \def \showISSN      #1{\unskip}     \fi
\ifx \showLCCN     \undefined \def \showLCCN      #1{\unskip}     \fi
\ifx \shownote     \undefined \def \shownote      #1{#1}          \fi
\ifx \showarticletitle \undefined \def \showarticletitle #1{#1}   \fi
\ifx \showURL      \undefined \def \showURL       {\relax}        \fi
\providecommand\bibfield[2]{#2}
\providecommand\bibinfo[2]{#2}
\providecommand\natexlab[1]{#1}
\providecommand\showeprint[2][]{arXiv:#2}

\bibitem[ACM(2020)]%
        {ACMBadges}
\bibfield{author}{\bibinfo{person}{ACM}.} \bibinfo{year}{2020}\natexlab{}.
\newblock \bibinfo{title}{{Artifact Review and Badging}}.
\newblock
  \bibinfo{howpublished}{\url{https://www.acm.org/publications/policies/artifact-review-badging}}.
\newblock
\newblock
\shownote{[Online; accessed 2024-01-25]}.


\bibitem[Ardis et~al\mbox{.}(2015)]%
        {securriculum2014}
\bibfield{author}{\bibinfo{person}{Mark Ardis}, \bibinfo{person}{David Budgen},
  \bibinfo{person}{Gregory~W. Hislop}, \bibinfo{person}{Jeff Offutt},
  \bibinfo{person}{Mark Sebern}, {and} \bibinfo{person}{Willem Visser}.}
  \bibinfo{year}{2015}\natexlab{}.
\newblock \showarticletitle{SE 2014: Curriculum Guidelines for Undergraduate
  Degree Programs in Software Engineering}.
\newblock \bibinfo{journal}{\emph{Computer}} \bibinfo{volume}{48},
  \bibinfo{number}{11} (\bibinfo{year}{2015}), \bibinfo{pages}{106--109}.
\newblock
\urldef\tempurl%
\url{https://doi.org/10.1109/MC.2015.345}
\showDOI{\tempurl}


\bibitem[Bartling and Friesike(2014)]%
        {Bartling2014}
\bibfield{author}{\bibinfo{person}{S{\"o}nke Bartling} {and}
  \bibinfo{person}{Sascha Friesike}.} \bibinfo{year}{2014}\natexlab{}.
\newblock \bibinfo{booktitle}{\emph{Towards Another Scientific Revolution}}.
\newblock \bibinfo{publisher}{Springer International Publishing},
  \bibinfo{address}{Cham}, \bibinfo{pages}{3--15}.
\newblock
\showISBNx{978-3-319-00026-8}
\urldef\tempurl%
\url{https://doi.org/10.1007/978-3-319-00026-8_1}
\showDOI{\tempurl}


\bibitem[Carpentries(2023)]%
        {TheCarpentries}
\bibfield{author}{\bibinfo{person}{The Carpentries}.}
  \bibinfo{year}{2023}\natexlab{}.
\newblock \bibinfo{title}{{The Carpentries}}.
\newblock \bibinfo{howpublished}{\url{https://carpentries.org/}}.
\newblock
\newblock
\shownote{[Online; accessed 2024-01-25]}.


\bibitem[Chawla(2021)]%
        {Chawla2021}
\bibfield{author}{\bibinfo{person}{Dalmeet~Singh Chawla}.}
  \bibinfo{year}{2021}\natexlab{}.
\newblock \bibinfo{title}{{Nature Index: ``Scientists at odds on Utrecht
  University reforms to hiring and promotion criteria''}}.
\newblock
  \bibinfo{howpublished}{\url{https://www.nature.com/nature-index/news/scientists-argue-over-use-of-impact-factors-for-evaluating-research}}.
\newblock
\newblock
\shownote{[Online; accessed 2024-01-25]}.


\bibitem[Chirigati et~al\mbox{.}(2016)]%
        {Chirigati:SIGMOD:2016}
\bibfield{author}{\bibinfo{person}{Fernando Chirigati},
  \bibinfo{person}{R\'{e}mi Rampin}, \bibinfo{person}{Dennis Shasha}, {and}
  \bibinfo{person}{Juliana Freire}.} \bibinfo{year}{2016}\natexlab{}.
\newblock \showarticletitle{ReproZip: Computational Reproducibility With Ease}.
  In \bibinfo{booktitle}{\emph{Proc. the 2016 International Conference on
  Management of Data (SIGMOD '16)}} (San Francisco, USA).
  \bibinfo{publisher}{ACM}, \bibinfo{address}{New York, NY, USA},
  \bibinfo{pages}{2085–2088}.
\newblock
\showISBNx{9781450335317}
\urldef\tempurl%
\url{https://doi.org/10.1145/2882903.2899401}
\showDOI{\tempurl}


\bibitem[Chue~Hong et~al\mbox{.}(2022)]%
        {chue_hong_fair_2022}
\bibfield{author}{\bibinfo{person}{Neil~P. Chue~Hong},
  \bibinfo{person}{Daniel~S. Katz}, \bibinfo{person}{Michelle Barker},
  {et~al\mbox{.}}} \bibinfo{year}{2022}\natexlab{}.
\newblock \bibinfo{title}{{FAIR Principles for Research Software (FAIR4RS
  Principles)}}.
\newblock
\newblock
\urldef\tempurl%
\url{https://doi.org/10.15497/RDA00068}
\showDOI{\tempurl}


\bibitem[Cohen et~al\mbox{.}(2021)]%
        {Cohen:IEEESoftware:2020}
\bibfield{author}{\bibinfo{person}{Jeremy Cohen}, \bibinfo{person}{Daniel~S.
  Katz}, \bibinfo{person}{Michelle Barker}, \bibinfo{person}{Neil Chue~Hong},
  \bibinfo{person}{Robert Haines}, {and} \bibinfo{person}{Caroline Jay}.}
  \bibinfo{year}{2021}\natexlab{}.
\newblock \showarticletitle{The Four Pillars of Research Software Engineering}.
\newblock \bibinfo{journal}{\emph{IEEE Software}} \bibinfo{volume}{38},
  \bibinfo{number}{1} (\bibinfo{year}{2021}), \bibinfo{pages}{97--105}.
\newblock
\urldef\tempurl%
\url{https://doi.org/10.1109/MS.2020.2973362}
\showDOI{\tempurl}


\bibitem[Commission et~al\mbox{.}(2017)]%
        {European2017}
\bibfield{author}{\bibinfo{person}{European Commission},
  \bibinfo{person}{Directorate-General for Research{ }and{ }Innovation},
  \bibinfo{person}{Cecili~Cabello Valdes}, \bibinfo{person}{Bernard Rentier},
  \bibinfo{person}{Eeva Kaunismaa}, {et~al\mbox{.}}}
  \bibinfo{year}{2017}\natexlab{}.
\newblock \bibinfo{booktitle}{\emph{Evaluation of research careers fully
  acknowledging Open Science practices: Rewards, incentives and/or recognition
  for researchers practicing Open Science}}.
\newblock \bibinfo{publisher}{Publications Office}, \bibinfo{address}{Europe}.
\newblock
\urldef\tempurl%
\url{https://doi.org/10.2777/75255}
\showDOI{\tempurl}


\bibitem[Consortium(2024)]%
        {rse:princ}
\bibfield{author}{\bibinfo{person}{RCP Consortium}.}
  \bibinfo{year}{2024}\natexlab{}.
\newblock \bibinfo{title}{{Princeton Research Computing -- Research Software
  Engineering}}.
\newblock
\newblock
\urldef\tempurl%
\url{https://researchcomputing.princeton.edu/services/research-software-engineering}
\showURL{%
\tempurl}
\newblock
\shownote{[Online; accessed 2024-01-27]}.


\bibitem[Cordeiro and OliveiraJr(2021)]%
        {CordeiroOliveiraJr2021}
\bibfield{author}{\bibinfo{person}{André Cordeiro} {and}
  \bibinfo{person}{Edson OliveiraJr}.} \bibinfo{year}{2021}\natexlab{}.
\newblock \showarticletitle{Open Science Practices for Software Engineering
  Controlled Experiments and Quasi-Experiments}. In
  \bibinfo{booktitle}{\emph{Proc. of the 1st Workshop on Open Science Practices
  for Software Engineering (OpenScienSE '21)}}. \bibinfo{publisher}{SBC},
  \bibinfo{address}{Porto Alegre, Brazil}, \bibinfo{pages}{19--21}.
\newblock
\urldef\tempurl%
\url{https://doi.org/10.5753/opensciense.2021.17140}
\showDOI{\tempurl}


\bibitem[Cordeiro(2022)]%
        {Cordeiro2022}
\bibfield{author}{\bibinfo{person}{Andr\'{e} Felipe~R. Cordeiro}.}
  \bibinfo{year}{2022}\natexlab{}.
\newblock \showarticletitle{An Open Science-Based Framework for Managing
  Experimental Data in Software Engineering}. In
  \bibinfo{booktitle}{\emph{Proc. of the 26th International Conference on
  Evaluation and Assessment in Software Engineering (EASE '22)}} (Gothenburg,
  Sweden). \bibinfo{publisher}{ACM}, \bibinfo{address}{New York, NY, USA},
  \bibinfo{pages}{342–346}.
\newblock
\showISBNx{9781450396134}
\urldef\tempurl%
\url{https://doi.org/10.1145/3530019.3535348}
\showDOI{\tempurl}


\bibitem[Durieux(2023)]%
        {AnonymousGithub}
\bibfield{author}{\bibinfo{person}{Thomas Durieux}.}
  \bibinfo{year}{2023}\natexlab{}.
\newblock \bibinfo{title}{{Anonymous GitHub}}.
\newblock \bibinfo{howpublished}{\url{https://anonymous.4open.science/}}.
\newblock
\newblock
\shownote{[Online; accessed 2024-01-25]}.


\bibitem[Ernst and Baldassarre(2023)]%
        {NeilBaldassarre2023}
\bibfield{author}{\bibinfo{person}{Neil~A. Ernst} {and}
  \bibinfo{person}{Maria~Teresa Baldassarre}.} \bibinfo{year}{2023}\natexlab{}.
\newblock \showarticletitle{Registered reports in software engineering}.
\newblock \bibinfo{journal}{\emph{Empirical Software Engineering}}
  \bibinfo{volume}{28}, \bibinfo{number}{2} (\bibinfo{date}{mar}
  \bibinfo{year}{2023}), \bibinfo{numpages}{11}~pages.
\newblock
\showISSN{1382-3256}
\urldef\tempurl%
\url{https://doi.org/10.1007/s10664-022-10277-5}
\showDOI{\tempurl}


\bibitem[FOSTER(2023)]%
        {FOSTER}
\bibfield{author}{\bibinfo{person}{FOSTER}.} \bibinfo{year}{2023}\natexlab{}.
\newblock \bibinfo{title}{{The FOSTER Portal: Open Science Training Courses}}.
\newblock
  \bibinfo{howpublished}{\url{https://www.fosteropenscience.eu/toolkit}}.
\newblock
\newblock
\shownote{[Online; accessed 2024-01-25]}.


\bibitem[Gonzalez-Barahona and Robles(2023)]%
        {Gonzalez-BarahonaRobles2023}
\bibfield{author}{\bibinfo{person}{Jesus~M. Gonzalez-Barahona} {and}
  \bibinfo{person}{Gregorio Robles}.} \bibinfo{year}{2023}\natexlab{}.
\newblock \showarticletitle{Revisiting the reproducibility of empirical
  software engineering studies based on data retrieved from development
  repositories}.
\newblock \bibinfo{journal}{\emph{Information and Software Technology}}
  \bibinfo{volume}{164} (\bibinfo{year}{2023}), \bibinfo{pages}{107318}.
\newblock
\showISSN{0950-5849}
\urldef\tempurl%
\url{https://doi.org/10.1016/j.infsof.2023.107318}
\showDOI{\tempurl}


\bibitem[Guzzo et~al\mbox{.}(2022)]%
        {Guzzo:IOP:2022}
\bibfield{author}{\bibinfo{person}{Richard~A. Guzzo}, \bibinfo{person}{Benjamin
  Schneider}, {and} \bibinfo{person}{Haig~R. Nalbantian}.}
  \bibinfo{year}{2022}\natexlab{}.
\newblock \showarticletitle{Open science, closed doors: The perils and
  potential of open science for research in practice}.
\newblock \bibinfo{journal}{\emph{Industrial and Organizational Psychology}}
  \bibinfo{volume}{15}, \bibinfo{number}{4} (\bibinfo{year}{2022}),
  \bibinfo{pages}{495–515}.
\newblock
\urldef\tempurl%
\url{https://doi.org/10.1017/iop.2022.61}
\showDOI{\tempurl}


\bibitem[Hettrick et~al\mbox{.}(2015)]%
        {hettrick2014uk}
\bibfield{author}{\bibinfo{person}{Simon Hettrick}, \bibinfo{person}{Mario
  Antonioletti}, \bibinfo{person}{Les Carr}, \bibinfo{person}{Neil Chue~Hong},
  \bibinfo{person}{Stephen Crouch}, \bibinfo{person}{David De~Roure},
  \bibinfo{person}{Iain Emsley}, \bibinfo{person}{Carole Goble},
  \bibinfo{person}{Alexander Hay}, \bibinfo{person}{Devasena Inupakutika},
  \bibinfo{person}{Mike Jackson}, \bibinfo{person}{Aleksandra Nenadic},
  \bibinfo{person}{Tim Parkinson}, \bibinfo{person}{Mark~I Parsons},
  \bibinfo{person}{Aleksandra Pawlik}, \bibinfo{person}{Giacomo Peru},
  \bibinfo{person}{Arno Proeme}, \bibinfo{person}{John Robinson}, {and}
  \bibinfo{person}{Shoaib Sufi}.} \bibinfo{year}{2015}\natexlab{}.
\newblock \bibinfo{booktitle}{\emph{UK Research Software Survey 2014}}.
\newblock
\urldef\tempurl%
\url{https://doi.org/10.5281/zenodo.14809}
\showDOI{\tempurl}


\bibitem[Howison and Herbsleb(2011)]%
        {howison2011scientific}
\bibfield{author}{\bibinfo{person}{James Howison} {and}
  \bibinfo{person}{James~D. Herbsleb}.} \bibinfo{year}{2011}\natexlab{}.
\newblock \showarticletitle{Scientific software production: incentives and
  collaboration}. In \bibinfo{booktitle}{\emph{Proc. of the ACM 2011 Conference
  on Computer Supported Cooperative Work (CSCW '11)}} (Hangzhou, China).
  \bibinfo{publisher}{Association for Computing Machinery},
  \bibinfo{address}{New York, NY, USA}, \bibinfo{pages}{513–522}.
\newblock
\showISBNx{9781450305563}
\urldef\tempurl%
\url{https://doi.org/10.1145/1958824.1958904}
\showDOI{\tempurl}


\bibitem[Howison and Herbsleb(2013)]%
        {howison2013incentives}
\bibfield{author}{\bibinfo{person}{James Howison} {and}
  \bibinfo{person}{James~D. Herbsleb}.} \bibinfo{year}{2013}\natexlab{}.
\newblock \showarticletitle{Incentives and integration in scientific software
  production}. In \bibinfo{booktitle}{\emph{Proc. of the 2013 Conference on
  Computer Supported Cooperative Work (CSCW '13)}} (San Antonio, Texas, USA).
  \bibinfo{publisher}{Association for Computing Machinery},
  \bibinfo{address}{New York, NY, USA}, \bibinfo{pages}{459–470}.
\newblock
\showISBNx{9781450313315}
\urldef\tempurl%
\url{https://doi.org/10.1145/2441776.2441828}
\showDOI{\tempurl}


\bibitem[Katz and Hettrick(2023)]%
        {KatzHettrick2023}
\bibfield{author}{\bibinfo{person}{D.~S. Katz} {and} \bibinfo{person}{S.
  Hettrick}.} \bibinfo{year}{2023}\natexlab{}.
\newblock \showarticletitle{Research Software Engineering in 2030}. In
  \bibinfo{booktitle}{\emph{2023 IEEE Conference on eScience}}.
  \bibinfo{publisher}{IEEE}, \bibinfo{address}{New York},
  \bibinfo{pages}{1--2}.
\newblock
\urldef\tempurl%
\url{https://doi.org/10.1109/e-Science58273.2023.10254813}
\showDOI{\tempurl}


\bibitem[Katz et~al\mbox{.}(2021)]%
        {katz:citation}
\bibfield{author}{\bibinfo{person}{Daniel~S. Katz}, \bibinfo{person}{Neil
  P.~Chue Hong}, \bibinfo{person}{Tim Clark}, {et~al\mbox{.}}}
  \bibinfo{year}{2021}\natexlab{}.
\newblock \showarticletitle{Recognizing the value of software: a software
  citation guide [version 2; peer review: 2 approved]}.
\newblock \bibinfo{journal}{\emph{F1000Research}} \bibinfo{volume}{9},
  \bibinfo{number}{1257} (\bibinfo{year}{2021}), \bibinfo{numpages}{13}~pages.
\newblock
\urldef\tempurl%
\url{https://doi.org/10.12688/f1000research.26932.2}
\showURL{%
\tempurl}


\bibitem[Levin et~al\mbox{.}(2016)]%
        {Levin_et_al2016}
\bibfield{author}{\bibinfo{person}{Nadine Levin}, \bibinfo{person}{Sabina
  Leonelli}, \bibinfo{person}{Dagmara Weckowska}, \bibinfo{person}{David
  Castle}, {and} \bibinfo{person}{John Dupré}.}
  \bibinfo{year}{2016}\natexlab{}.
\newblock \showarticletitle{How Do Scientists Define Openness? Exploring the
  Relationship Between Open Science Policies and Research Practice}.
\newblock \bibinfo{journal}{\emph{Bulletin of Science, Technology \& Society}}
  \bibinfo{volume}{36}, \bibinfo{number}{2} (\bibinfo{year}{2016}),
  \bibinfo{pages}{128--141}.
\newblock
\urldef\tempurl%
\url{https://doi.org/10.1177/0270467616668760}
\showDOI{\tempurl}


\bibitem[Mendez et~al\mbox{.}(2020)]%
        {mendez2020}
\bibfield{author}{\bibinfo{person}{Daniel Mendez}, \bibinfo{person}{Daniel
  Graziotin}, \bibinfo{person}{Stefan Wagner}, {and} \bibinfo{person}{Heidi
  Seibold}.} \bibinfo{year}{2020}\natexlab{}.
\newblock \bibinfo{booktitle}{\emph{Open Science in Software Engineering}}.
\newblock \bibinfo{publisher}{Springer International Publishing},
  \bibinfo{address}{Cham}, \bibinfo{pages}{477--501}.
\newblock
\showISBNx{978-3-030-32489-6}
\urldef\tempurl%
\url{https://doi.org/10.1007/978-3-030-32489-6_17}
\showDOI{\tempurl}


\bibitem[Munafò et~al\mbox{.}(2017)]%
        {munafo2017manifesto}
\bibfield{author}{\bibinfo{person}{Marcus~R. Munafò},
  \bibinfo{person}{Brian~A. Nosek}, \bibinfo{person}{Dorothy V.~M. Bishop},
  {et~al\mbox{.}}} \bibinfo{year}{2017}\natexlab{}.
\newblock \showarticletitle{A manifesto for reproducible science}.
\newblock \bibinfo{journal}{\emph{{Nature Human Behaviour}}}
  \bibinfo{volume}{1}, \bibinfo{number}{1} (\bibinfo{year}{2017}),
  \bibinfo{pages}{1--9}.
\newblock
\urldef\tempurl%
\url{https://doi.org/10.1038/s41562-016-0021}
\showDOI{\tempurl}


\bibitem[{National Academies of Sciences, Engineering, and Medicine}(2018)]%
        {nas2018}
\bibfield{author}{\bibinfo{person}{{National Academies of Sciences,
  Engineering, and Medicine}}.} \bibinfo{year}{2018}\natexlab{}.
\newblock \bibinfo{booktitle}{\emph{{Open Science by Design: Realizing a Vision
  for 21st Century Research}}}.
\newblock \bibinfo{publisher}{The National Academies Press},
  \bibinfo{address}{Washington, USA}.
\newblock
\urldef\tempurl%
\url{https://doi.org/10.17226/25116}
\showDOI{\tempurl}


\bibitem[OliveiraJr et~al\mbox{.}(2021)]%
        {OliveiraJr_et_al2021}
\bibfield{author}{\bibinfo{person}{Edson OliveiraJr}, \bibinfo{person}{Viviane
  Furtado}, \bibinfo{person}{Henrique Vignando}, \bibinfo{person}{Carlos Luz},
  \bibinfo{person}{Andr\'{e} Cordeiro}, \bibinfo{person}{Igor Steinmacher},
  {and} \bibinfo{person}{Avelino Zorzo}.} \bibinfo{year}{2021}\natexlab{}.
\newblock \showarticletitle{Towards Improving Experimentation in Software
  Engineering}. In \bibinfo{booktitle}{\emph{Proc. of the XXXV Brazilian
  Symposium on Software Engineering (SBES '21)}} (Joinville, Brazil).
  \bibinfo{publisher}{Association for Computing Machinery},
  \bibinfo{address}{New York, NY, USA}, \bibinfo{pages}{335–340}.
\newblock
\showISBNx{9781450390613}
\urldef\tempurl%
\url{https://doi.org/10.1145/3474624.3477073}
\showDOI{\tempurl}


\bibitem[OpenAire(2024)]%
        {OpenAire2024}
\bibfield{author}{\bibinfo{person}{OpenAire}.} \bibinfo{year}{2024}\natexlab{}.
\newblock \bibinfo{title}{{OpenAire - Open Science in Europe}}.
\newblock
\newblock
\urldef\tempurl%
\url{https://www.openaire.eu/projects}
\showURL{%
\tempurl}
\newblock
\shownote{[Online; accessed 2024-01-25]}.


\bibitem[{OpenSciency Contributors}(2023)]%
        {Opensciency2023}
\bibfield{author}{\bibinfo{person}{{OpenSciency Contributors}}.}
  \bibinfo{year}{2023}\natexlab{}.
\newblock \bibinfo{title}{{Opensciency - A core open science curriculum by and
  for the research community. Zenodo.}}
\newblock
\newblock
\urldef\tempurl%
\url{https://doi.org/10.5281/zenodo.7662732}
\showDOI{\tempurl}


\bibitem[Paul et~al\mbox{.}(2023)]%
        {Paul_et_al2023}
\bibfield{author}{\bibinfo{person}{Showmick~Guha Paul}, \bibinfo{person}{Arpa
  Saha}, \bibinfo{person}{Mohammad~Shamsul Arefin}, \bibinfo{person}{Touhid
  Bhuiyan}, \bibinfo{person}{Al~Amin Biswas}, \bibinfo{person}{Ahmed~Wasif
  Reza}, \bibinfo{person}{Naif~M. Alotaibi}, \bibinfo{person}{Salem~A. Alyami},
  {and} \bibinfo{person}{Mohammad~Ali Moni}.} \bibinfo{year}{2023}\natexlab{}.
\newblock \showarticletitle{A Comprehensive Review of Green Computing: Past,
  Present, and Future Research}.
\newblock \bibinfo{journal}{\emph{IEEE Access}}  \bibinfo{volume}{11}
  (\bibinfo{year}{2023}), \bibinfo{pages}{87445--87494}.
\newblock
\urldef\tempurl%
\url{https://doi.org/10.1109/ACCESS.2023.3304332}
\showDOI{\tempurl}


\bibitem[Pimentel et~al\mbox{.}(2017)]%
        {Pimentel:VLDB:2017}
\bibfield{author}{\bibinfo{person}{Jo\~{a}o~Felipe Pimentel},
  \bibinfo{person}{Leonardo Murta}, \bibinfo{person}{Vanessa Braganholo}, {and}
  \bibinfo{person}{Juliana Freire}.} \bibinfo{year}{2017}\natexlab{}.
\newblock \showarticletitle{noWorkflow: a tool for collecting, analyzing, and
  managing provenance from python scripts}.
\newblock \bibinfo{journal}{\emph{Proc. of the VLDB Endowment}}
  \bibinfo{volume}{10}, \bibinfo{number}{12} (\bibinfo{date}{aug}
  \bibinfo{year}{2017}), \bibinfo{pages}{1841–1844}.
\newblock
\showISSN{2150-8097}
\urldef\tempurl%
\url{https://doi.org/10.14778/3137765.3137789}
\showDOI{\tempurl}


\bibitem[Rampin et~al\mbox{.}(2018)]%
        {Rampin:2018:reproserver}
\bibfield{author}{\bibinfo{person}{Remi Rampin}, \bibinfo{person}{Fernando
  Chirigati}, \bibinfo{person}{Vicky Steeves}, {and} \bibinfo{person}{Juliana
  Freire}.} \bibinfo{year}{2018}\natexlab{}.
\newblock \bibinfo{title}{ReproServer: Making Reproducibility Easier and Less
  Intensive}.
\newblock
\newblock
\showeprint[arxiv]{1808.01406}~[cs.SE]


\bibitem[RSE.org(2024)]%
        {rse:org}
\bibfield{author}{\bibinfo{person}{RSE.org}.} \bibinfo{year}{2024}\natexlab{}.
\newblock \bibinfo{title}{{Society of Research Software Engineering}}.
\newblock
\newblock
\urldef\tempurl%
\url{https://society-rse.org}
\showURL{%
\tempurl}
\newblock
\shownote{[Online; accessed 2024-01-27]}.


\bibitem[RSE.Shef(2024)]%
        {rse:shef}
\bibfield{author}{\bibinfo{person}{RSE.Shef}.} \bibinfo{year}{2024}\natexlab{}.
\newblock \bibinfo{title}{{Research Software Engineering Sheffield}}.
\newblock
\newblock
\urldef\tempurl%
\url{https://rse.shef.ac.uk}
\showURL{%
\tempurl}
\newblock
\shownote{[Online; accessed 2024-01-27]}.


\bibitem[Santana et~al\mbox{.}(2023)]%
        {Santana_et_al2023}
\bibfield{author}{\bibinfo{person}{Filipe Santana}, \bibinfo{person}{André
  Cordeiro}, {and} \bibinfo{person}{Edson OliveiraJr}.}
  \bibinfo{year}{2023}\natexlab{}.
\newblock \showarticletitle{Use of the Dublin Core Standard to Express Open
  Metadata Related to Software Engineering Experiments}. In
  \bibinfo{booktitle}{\emph{Proc. of the 3rd Workshop on Open Science Practices
  for Software Engineering (OpenScienSE '23)}}. \bibinfo{publisher}{SBC},
  \bibinfo{address}{Porto Alegre, Brazil}, \bibinfo{pages}{1--5}.
\newblock
\showISSN{0000-0000}
\urldef\tempurl%
\url{https://doi.org/10.5753/opensciense.2023.235672}
\showDOI{\tempurl}


\bibitem[Santos(2021)]%
        {Santos:OpenScienSE:2021}
\bibfield{author}{\bibinfo{person}{Alcemir Santos}.}
  \bibinfo{year}{2021}\natexlab{}.
\newblock \showarticletitle{Open scientist in the wonderland: advocating for
  blockchain-based decentralized applications for science}. In
  \bibinfo{booktitle}{\emph{{Proc. of the 1st Workshop on Open Science
  Practices for Software Engineering (OpenScienSE '21)}}}.
  \bibinfo{publisher}{SBC}, \bibinfo{address}{Porto Alegre, Brazil},
  \bibinfo{pages}{34--36}.
\newblock
\showISSN{0000-0000}
\urldef\tempurl%
\url{https://doi.org/10.5753/opensciense.2021.17143}
\showDOI{\tempurl}


\bibitem[Seibold(2024)]%
        {Seibold2024}
\bibfield{author}{\bibinfo{person}{Heidi Seibold}.}
  \bibinfo{year}{2024}\natexlab{}.
\newblock \bibinfo{title}{{A}voiding fraud and improving rigor through {O}pen
  {S}cience}.
\newblock
\newblock
\urldef\tempurl%
\url{https://doi.org/10.34734/FZJ-2024-00813}
\showDOI{\tempurl}
\newblock
\shownote{Blog post}.


\bibitem[United Nations~Educational and (UNESCO)(2021)]%
        {unesco:2021}
\bibfield{author}{\bibinfo{person}{Scientific United Nations~Educational} {and}
  \bibinfo{person}{Cultural~Organization (UNESCO)}.}
  \bibinfo{year}{2021}\natexlab{}.
\newblock \bibinfo{title}{{UNESCO Recommendation on Open Science}}.
\newblock
\newblock
\urldef\tempurl%
\url{https://doi.org/10.5281/zenodo.5834767}
\showDOI{\tempurl}


\bibitem[{US-RSE.org}(2022)]%
        {rs:et2022}
\bibfield{author}{\bibinfo{person}{{US-RSE.org}}.}
  \bibinfo{year}{2022}\natexlab{}.
\newblock \bibinfo{title}{{Education and Training for Research Software
  Engineers}}.
\newblock
\newblock
\urldef\tempurl%
\url{https://us-rse.org/2022-08-01-education-training/}
\showURL{%
\tempurl}
\newblock
\shownote{[Online; accessed 2024-01-25]}.


\bibitem[Wiese et~al\mbox{.}(2020)]%
        {wiese:2020}
\bibfield{author}{\bibinfo{person}{Igor Wiese}, \bibinfo{person}{Ivanilton
  Polato}, {and} \bibinfo{person}{Gustavo Pinto}.}
  \bibinfo{year}{2020}\natexlab{}.
\newblock \showarticletitle{Naming the Pain in Developing Scientific Software}.
\newblock \bibinfo{journal}{\emph{IEEE Software}} \bibinfo{volume}{37},
  \bibinfo{number}{4} (\bibinfo{year}{2020}), \bibinfo{pages}{75--82}.
\newblock
\urldef\tempurl%
\url{https://doi.org/10.1109/MS.2019.2899838}
\showDOI{\tempurl}


\bibitem[Wilkinson et~al\mbox{.}(2016)]%
        {WilkinsonEtAl2016}
\bibfield{author}{\bibinfo{person}{Mark~D. Wilkinson}, \bibinfo{person}{Michel
  Dumontier}, \bibinfo{person}{IJsbrand~Jan Aalbersberg}, {et~al\mbox{.}}}
  \bibinfo{year}{2016}\natexlab{}.
\newblock \showarticletitle{{The FAIR Guiding Principles for Scientific Data
  Management and Stewardship}}.
\newblock \bibinfo{journal}{\emph{Scientific Data}}  \bibinfo{volume}{3}
  (\bibinfo{year}{2016}), \bibinfo{numpages}{9}~pages.
\newblock
\showISSN{2052-4463}
\urldef\tempurl%
\url{https://doi.org/10.1038/sdata.2016.18}
\showDOI{\tempurl}


\bibitem[{WOSSS}(2024)]%
        {rs:wosss}
\bibfield{author}{\bibinfo{person}{{WOSSS}}.} \bibinfo{year}{2024}\natexlab{}.
\newblock \bibinfo{title}{{Workshop on Sustainable Software Sustainability
  (WoSSS)}}.
\newblock
\newblock
\urldef\tempurl%
\url{https://wosss.org}
\showURL{%
\tempurl}
\newblock
\shownote{[Online; accessed 2024-01-25]}.


\end{thebibliography}

\end{document}